\documentclass[useAMS, usenatbib]{mn2e}
\usepackage{graphicx, rotate}
\usepackage{natbib, lscape}
\usepackage{latexsym, amssymb, verbatim}
\usepackage[toc,page,header]{appendix}

\title[Chemical tracers of high-metallicity environments]
{Chemical tracers of high-metallicity environments}
\author[Bayet et al.]{E. Bayet$^{1}$\thanks{E-mail:
bayet@physics.ox.ac.uk;}; T. A. Davis$^{2}$; T. A. Bell$^{3}$ and S. Viti$^{4}$\\
$^{1}$Sub-Department of astrophysics, University of Oxford, Denys
Wilkinson Building, Keble Road, Oxford OX1 3RH, UK\\
$^{2}$European Southern Observatory, Karl-Schwarzschild-Str. 2,
85748 Garching, Germany\\
$^{3}$Centro de Astrobiolog\'ia (CSIC-INTA), Carretera de Ajalvir km 4, 28850 Madrid, Spain\\
$^{4}$Department of Physics and Astronomy, University
College London, Gower Street, London WC1E 6BT, UK\\
}

\begin{document}

\date{Accepted ; Received ; in original form }

\pagerange{\pageref{firstpage}--\pageref{lastpage}} \pubyear{2010}

\maketitle

\label{firstpage}

\begin{abstract}
We present for the first time a detailed study of the properties
of molecular gas in metal-rich environments such as early-type
galaxies (ETGs). We have explored Photon-Dominated Region (PDR)
chemistry for a wide range of physical conditions likely to be
appropriate for these sources. We derive fractional abundances of
the 20 most chemically reactive species as a function of the
metallicity, as a function of the optical depth and for various
volume number gas densities, Far-Ultra Violet (FUV) radiation
fields and cosmic ray ionisation rates. We also investigate the
response of the chemistry to the changes in $\alpha-$element
enhancement as seen in ETGs. We find that the fractional
abundances of CS, H$_{2}$S, H$_{2}$CS, H$_{2}$O, H$_{3}$O$^{+}$,
HCO$^{+}$ and H$_{2}$CN seem invariant to an increase of
metallicity whereas C$^{+}$, CO, C$_{2}$H, CN, HCN, HNC and OCS
appear to be the species most sensitive to this change. The most
sensitive species to the change in the fractional abundance of
$\alpha-$elements are C$^{+}$, C, CN, HCN, HNC, SO, SO$_{2}$,
H$_{2}$O and CS. Finally, we provide line brightness ratios for
the most abundant species, especially in the range observable with
ALMA. Discussion of favorable line ratios to use for the
estimation of super-solar metallicities and $\alpha$-elements are
also provided.
\end{abstract}


\section{Introduction}\label{sec:intro}

Understanding the production of metals, and the subsequent
enrichment of the ISM is crucial in many field of extragalactic
astrophysics. For example, the initial mass function (IMF) of
stars must have been different in the early universe, where no
metals were present, and whether there is any remaining variation
of IMF with metallicity remains an open question (e.g.
\citealt{Omuk05,Santo06,Tuml07,Vando11}). Gas metallicity is
regulated by a complex interplay between star formation, infall of
metal-poor gas and outflow of enriched material. The discovery of
a relation between galaxy mass and metallicity \citep{Lequ79b},
showing that the more massive a galaxy is the higher is its
metallicity, has increased debate about the metallicity and its
relationship with galaxy characteristics and in particular with
the properties of its interstellar medium.

Estimating the metallicity of a galaxy remains to date a difficult
task with few methods available, and these have large
uncertainties. Stellar metallicities are usually derived from
absorption indices (such as those of magnesium and iron), but
these weak features are hard to observe at high redshift, and only
give information about the metallicity of the system in the past,
when the stars were formed (e.g. \citealt{McDe11}). Optically
determined metallicities of the ionised gas provide information on
the metal content of star-forming complexes, allowing astronomers
to probe the gas phase metallicity further out into the universe
but this approach only works if the emission spectrum observed is
powered by star formation processes. Where strong AGN activity,
old stellar populations or shocks dominate, this method becomes
unreliable. Furthermore, there can be strong degeneracies between
low and high metallicity populations in some ionised gas
metallicity indicators.

In this paper, we explore the possibility of determining
metallicity via submillimeter observations where no dust
attenuation is present, remaining a reliable method even if star
formation does not dominate the ionisation of the gas. In
particular, we concentrate on exploring the high metallicity range
represented by early-type galaxies (hereafter called ETGs). These
metal-rich objects (ellipticals and lenticulars) located at the
`end point' of galaxy merger sequences, carry important signatures
of mass assembly and star formation in the Universe (e.g.
\citealt{Kavi09}). Surprisingly, a significant fraction of them
($\sim$ 20\%) contain some molecular gas reservoirs and are not as
`red-and-dead' as previously thought \citep{Welc03, Sage06,
Comb07, Sage07, Krip10, Youn11, Croc11}. The properties of this
molecular gas such as its temperature, volume number gas density,
column density of gas etc. have been estimated \citep{Baye12a}
showing a priori similarity with the molecular gas properties seen
in the center of our own Galaxy. However, the origin of the gas in
ETGs remains to date a mystery (but see \citealt{Davi11a}).
Determining the best molecular gas metallicity tracer can help to
better constrain this origin. Indeed, if the metallicity of the
molecular gas is similar to the measurements obtained from stars
and the ionised gas, then the origin of the gas in ETGs is likely
to be internal due to stellar mass loss. Otherwise, it is argued
that an external origin due to past merging events and/or
accretion is more likely.

Here, we perform a first detailed and comprehensive
theoretical study of the influence of various parameters related
to metallicity on the molecular gas properties, all likely to be
appropriate for ETGs. The Photon-Dominated Region (PDR) code and
the chemical database that we have used for our study are
described in Section \ref{sec:mod}, and the physical parameters we
have selected are discussed in Section \ref{sec:param}. Results
describing the chemistry and its sensitivity to variations of
these parameters are presented in Section \ref{sec:resu}. In
Section \ref{sec:ana}, we list the best chemical species or
combinations of them we found which represent well ETG
environments. In Section \ref{sec:con} we summarize our
conclusions.

\section{Model Description}\label{sec:mod}

We make use of the UCL\_PDR model as described in \citet{Bell05,
Bell06, Baye09a, Baye11b}. This is a time-dependent
Photon-Dominated Region (PDR) model. In the present application,
the code is run for 10$^{7}$ yrs, at constant density, by which
time chemical steady state is reached in all models. In fact, the
fractional abundances at 10$^{6}$ yrs do not differ significantly
from those at 10$^{7}$ yrs. Thus, time-dependence above 10$^{6}$
yrs is unlikely to play a role in the chemistry. The code operates
in one spatial dimension for an assumed semi-infinite slab
geometry, and computes self-consistently the chemistry and the
temperature as functions of depth and time within the
semi-infinite slab, taking account of a wide range of heating and
cooling processes \citep{Baye11b}. In this work, the code is used
to determine the chemical and thermal properties at all depths up
to 20 visual magnitudes.

Here, we improved the UCL\_PDR code by adding the calculation of
the line brightnesses for $^{13}$CO, C$^{18}$O, CS and HCN
transitions (see results in Subsect \ref{sec:resu}). These line
brightnesses (in erg s$^{-1}$ cm$^{-2}$ sr$^{-1}$) are calculated
using the LVG approximation (see e.g. \citealt{VanderTak07}) and
the collisional rates with H$_2$ are taken from the Leiden Atomic
and Molecular Database (LAMDA; \citealt{Scho05}). The detailed
description of these latest changes in the code and their
consequences in its overall structure are outside the scope of
this paper but can be found in great detail in \citet{Bell12}.

The chemical network used here links 131 species in over 1800
gas-phase reactions; only H$_{2}$ is formed by surface chemistry.
Freeze-out of species on to grain surfaces is not considered. This
might have a severe impact on the fractional abundance predictions
at A$_{\rm v}>3-8$ mag since at these optical depths, the
temperatures, as seen in Fig. \ref{fig:1}, are low. In addition,
freeze-out time is about 10$^{4}$yrs for most of the models
developed in the present study and 1000 yrs and about 100 yrs for
Models 13 and 14 respectively. Therefore, freeze-out may be
important. However, at those extinctions the rotational lines
considered in this study are mostly optically thick in which case
freeze-out would not affect the line ratios but still the
abundances. We have thus run chemical models to estimate the
impact of freeze-out on abundances and this is estimated to be
about 3-8\%, depending on the extinction and species considered.
The UCL\_PDR code has been validated against all other commonly
used PDR codes and performs well \citep{Roel07}.

\section{Parameter Selection}\label{sec:param}

The model requires the setting of a number of physical and
chemical parameters. The choices we have made are listed in Table
\ref{tab:1}, \ref{tab:2} and \ref{tab:3}.

Most of the calculations are made for an assumed gas number
density of $1\times 10^{4}$ hydrogen nuclei (in all forms)
cm$^{-3}$, typical value found for molecular gas in galaxies. The
A$_{\rm v}$ values range from 0 to 20 mags. For Fig. \ref{fig:1}
(left hand side) and Fig. \ref{fig:2}, we took five representative
depths into the one-dimensional slab: A$_{\rm v}=1$, 3, 5, 8 and
20 mag. These values have been selected intending to represent
various gas phases from a dense PDR surface to dark cloud
conditions.

We have investigated how metallicity changes influence the
chemistry by studying the effect of an increase of metallicity
from 1 to 3.3 z$_{\odot}$ by steps of 0.2 z$_{\odot}$ from 1.1
z$_{\odot}$. z = 1 z$_{\odot}$ corresponds to solar value of the
initial elemental abundance ratios (see Table \ref{tab:2}) while z
= 2.5 z$_{\odot}$ means that the solar values of the initial
elemental abundance ratios have been all multiplied by the same
factor (of 2.5 in this example). The metallicity range has been
chosen such as to agree with observational results from the
ATLAS$^{\rm 3D}$ survey which show that metallicity ranges for
local ETGs from 0.043 to 0.5 dex i.e. from $~$ 0.9 z$_{\odot}$ to
$~$ 3.2 z$_{\odot}$ \citep{McDe11}. The step of 0.2 z$_{\odot}$
was chosen so as to have a reasonable sampling. Several parameters
are assumed to scale linearly with metallicity. These are the
dust-to-gas mass ratio, the H$_{2}$ formation rate, and the
initial elemental abundances. The values of these parameters for
solar metallicity are given in Table \ref{tab:1} and \ref{tab:2}.
The elemental initial abundances given in Table \ref{tab:2} are
derived from the most up-to-date estimates obtained on comets and
solar photosphere studies (see \citealt{Aspl09} and references
therein).

High metallicity environments such as ETGs or Milky Way metal-rich
stars (typically Pop II) have been found to have some of their
elements enhanced as compared to others (e.g. \citealt{Ryan91,
McDe11}). These enhanced elements are oxygen (O), neon (Ne),
magnesium (Mg), silicon (Si), sulphur (S), argon (Ar), calcium
(Ca) and titanium (Ti). They are more generally called
$\alpha-$elements since their most abundant isotopes are integer
multiples of the mass of the helium nucleus (the $\alpha$
particle). \citet{All60} were the first to see $\alpha-$element
enhancement in metal-poor stars in our Galaxy. \citet{Pele89} was
the first to note that you can not reproduce the stellar
populations of ETGs without $\alpha-$element enhancement whereas
\citet{Rana03, Orig04} worked out what this meant in terms of
bursty star formation. For a review on $\alpha-$elements and their
enhancements, we refer the reader to \citet{Thom99}.

In the current UCL\_PDR code, only 5/8 $\alpha-$elements, namely
O, Mg, Si, S and Ca are included in the chemical network (see
Table \ref{tab:2} in bold font). For the rest of the paper, we
thus consider only the influence on the chemistry of a change in
these 5 $\alpha-$elements (see Subsection \ref{subsec:alpha}). To
be more specific, we studied the $\alpha-$element enhancement of
\citet{Sala98} who provided the most recent $\alpha-$elements
enhanced mixture able to reproduce well observations of the
metal-rich globular clusters in the galactic disk (see Table
\ref{tab:2}, column 2) and thought to be appropriate for ETGs. In
their study, they enhanced the $\alpha-$elements by various
factors. We also studied another scenario of $\alpha-$element
enhancement used mostly in metal-rich stellar evolution model
likely appropriate for ETGs too, investigated by \citet{Weis04}.
In their paper, contrarily to \citet{Sala98}, they applied a
systematic +0.4 dex factor to all the $\alpha-$elements. Here, we
do not use directly the values reported in \citet{Sala98} and
\citet{Weis04} because they each use different standard (solar)
values than ours but we adapted their approach to our standard
values (see results in Table \ref{tab:2}).

The models require the specification of the mean dust grain radius
and albedo, which affect the transfer of external radiation into
the cloud. The values given in Table \ref{tab:1} are canonical.
The external radiation field impinging on the one-dimensional
semi-infinite slab is assumed to be 1 Habing in most cases,
however the influence of stronger radiation field is investigated
through Models 15 and 16 (see Section \ref{sec:resu}). 1 Habing is
the standard value used for the Milky Way. The cosmic ray
ionisation rate standard values used in our study is $5 \times
10^{-17}$ s$^{-1}$.

Finally, the models also require the specification of the
microturbulence velocity for line-width calculations . We use the
value adopted by \citet{Baye09a} i.e. 1.5 km s$^{-1}$.

Our grid contains 20 models. Models 0 to 12 investigate the effect
of metallicity changes on the chemistry. Models 8, 13 and 14 allow
us to study the influence on the chemistry of the change of gas
number density. Models 15 and 16 look at the influence of an
enhanced FUV radiation field, whereas Models 17 and 18 focus on
the influence of the increase of cosmic ray ionisation rate.
Finally Models 19 and 20 analyse changes in the high-metallicity
chemistry for different scenarios of $\alpha-$element enhancement.

As shown in \citet{McDe11} and \citet{Kunt10}, some ETGs may have
sub-solar metallicity stellar populations in their cores. These
low metallicity cores are likely caused by the accretion of low
metallicity gas. \citet{Kavi11} showed that a number of ETGs have
acquired cold gas through accretion during minor merger events
with dwarf galaxies (at redshift smaller than 1). \citet{Davi11b}
also showed that $>$50\% of gas-rich field ETGs in the local
volume have obtained this gas from mergers or accretion. Low
metallicity gas that is accreted will, however, quickly be
enriched by SNII (within few hundred Myrs) from the newly formed
stellar population. It is thus likely that almost all gas rich
ETGs have solar metallicity gas, or above. This assumption is
reinforced by the fact that in cluster environments, where most of
the ETGs studied in this paper are, the dwarf satellite galaxies
are coming in fairly dry, having most of its gas already stripped
out, and so, a larger fraction of the gas is actually coming from
stellar mass loss and therefore must be of higher metallicity. The
chemistry of low metallicity gas has been partially explored in
\citet{Baye09a} and forthcoming more detailed studies are in
preparation.

\begin{table*}
    \caption{Standard model parameters (see text in Sect. \ref{sec:param}).
    These parameters are those used in Models 0-20 except if a different
    value is specified in Table \ref{tab:3} or described in the text.}\label{tab:1}
    \begin{tabular}{l r}
    \hline
    Gas Density & 10$^{4}$ cm$^{-3}$\\
    Mean photon scattering by grain & 0.9\\
    External FUV radiation intensity (=I$_{\odot}$) & 1 Habing$^{a}$ \\
    Cosmic ray ionization rate (=$\zeta_{\odot}$) & 5.0$\times 10^{-17}$ s$^{-1}$\\
    H$_{2}$ formation rate coefficient (=R$_{\odot}$) & 3$\times 10^{-18}\sqrt{T} \exp(\frac{-T}{1000})$ cm$^{3}$s$^{-1}$\\
    Dust:gas mass ratio (=d$_{\odot}$) & 1/100 \\
    Metallicity (=z$_{\odot}$) & solar values $^{b}$\\
    Grain size & 0.1 $\micron$\\
    Grain albedo & 0.7 \\
    A$_{\rm v}$ maximum & 20 mag\\
    Microturbulent velocity & 1.5 kms$^{-1}$\\
    \hline
    \end{tabular}

    $^{a}$: The unit of the standard  Interstellar Radiation Field (ISRF) intensity
    is I$_{\odot}$ = 1.6$\times 10^{-3}$ erg cm$^{-2}$ s$^{-1}$ \citet{Habi68};
    $^{b}$ : z = 1 = z$_{\odot}$ corresponds to solar values of the initial
    elemental abundance ratios (see Table \ref{tab:2}, first column) while
    z = 2.5 z$_{\odot}$ means that the solar values of the initial elemental
    abundance ratios have been all multiplied by the same factor (of 2.5 in this
    example).
\end{table*}

\begin{table}
    \caption{Initial abundance ratios used in Table \ref{tab:3}. The abbreviations
    `SW98' and `WE04' refer to \citet{Sala98} and \citet{Weis04}, respectively
    (see Sect. \ref{sec:param}). The standard initial elemental abundance ratios
    (`ST') are from \citet{Aspl09}. Models using the `ST' values of the initial
    elemental abundance ratios are the models corresponding to z = 1 z$_{\odot}$.
    Bolt font is used to represent the $\alpha-$elements.}\label{tab:2}
    \begin{center}
    \begin{tabular}{c c c c}
    \hline
    & ST & SW98 & WE04 \\
    \hline
    Fe/H & 3.16$\times$10$^{-5}$ & 3.16$\times$10$^{-5}$ & 3.16$\times$10$^{-5}$\\
    C/H  & 2.69$\times$10$^{-4}$ & 2.69$\times$10$^{-4}$ & 2.69$\times$10$^{-4}$ \\
    \textbf{O/H}  & 4.90$\times$10$^{-4}$ & 1.55$\times$10$^{-3}$ & 1.23$\times$10$^{-3}$ \\
    N/H  & 6.76$\times$10$^{-5}$ & 6.76$\times$10$^{-5}$ & 6.76$\times$10$^{-5}$\\
    \textbf{Si/H} & 3.24$\times$10$^{-5}$ & 8.13$\times$10$^{-5}$ & 8.13$\times$10$^{-5}$\\
    \textbf{S/H}  & 1.32$\times$10$^{-5}$ & 2.82$\times$10$^{-5}$ & 3.31$\times$10$^{-5}$ \\
    \textbf{Ca/H} & 2.19$\times$10$^{-6}$ & 6.92$\times$10$^{-6}$ & 5.50$\times$10$^{-6}$\\
    He/H & 8.51$\times$10$^{-2}$ & 8.51$\times$10$^{-2}$ & 8.51$\times$10$^{-2}$ \\
    \textbf{Mg/H} & 3.98$\times$10$^{-5}$ & 1.00$\times$10$^{-4}$ & 1.00$\times$10$^{-4}$\\
    Na/H & 1.74$\times$10$^{-6}$ & 1.74$\times$10$^{-6}$ & 1.74$\times$10$^{-6}$\\
    Cl/H & 3.16$\times$10$^{-7}$ & 3.16$\times$10$^{-7}$ & 3.16$\times$10$^{-7}$\\
    \hline
    \end{tabular}
    \end{center}
\end{table}

\begin{table*}
    \caption{Input parameters of the UCL\_PDR models used to
    perform this study (see Sections \ref{sec:param} and \ref{sec:resu}).
    The abbreviation `ST' represents the standard values of the initial
    elemental abundance ratios while
    the abbreviations `SW98' and `WE04' correspond to the values of initial
    elemental abundance ratios referenced in Table
    \ref{tab:2}. This table does not present all the
    input parameters of the UCL\_PDR code for each model but only
    lists the parameters set to values different from the standard
    ones (see also Table \ref{tab:1} for complementary information).}\label{tab:3}
    \begin{center}
    \begin{tabular}{r c c c c c c c c}
    \hline
    Model & Metallicity   & Dust-to-gas              & H$_{2}$ form.            & Ini. Elem.    & FUV rad. field & $\zeta$           & Gas Density\\
          & (z$_{\odot}$) & mass ratio (d$_{\odot}$) & rate coeff.(R$_{\odot}$) & Abund. ratios & (I$_{\odot}$)  & ($\zeta_{\odot}$) & (cm$^{-3}$)\\
    \hline
    0     & 1             & 1                        & 1                        & ST            & 1              & 1                 & 10$^{4}$\\
    1     & 1.1           & 1.1                      & 1.1                      & STx1.1        & 1              & 1                 & 10$^{4}$\\
    2     & 1.3           & 1.3                      & 1.3                      & STx1.3        & 1              & 1                 & 10$^{4}$\\
    3     & 1.5           & 1.5                      & 1.5                      & STx1.5        & 1              & 1                 & 10$^{4}$\\
    4     & 1.7           & 1.7                      & 1.7                      & STx1.7        & 1              & 1                 & 10$^{4}$\\
    5     & 1.9           & 1.9                      & 1.9                      & STx1.9        & 1              & 1                 & 10$^{4}$\\
    6     & 2.1           & 2.1                      & 2.1                      & STx2.1        & 1              & 1                 & 10$^{4}$\\
    7     & 2.3           & 2.3                      & 2.3                      & STx2.3        & 1              & 1                 & 10$^{4}$\\
    8     & 2.5           & 2.5                      & 2.5                      & STx2.5        & 1              & 1                 & 10$^{4}$\\
    9     & 2.7           & 2.7                      & 2.7                      & STx2.7        & 1              & 1                 & 10$^{4}$\\
   10     & 2.9           & 2.9                      & 2.9                      & STx2.9        & 1              & 1                 & 10$^{4}$\\
   11     & 3.1           & 3.1                      & 3.1                      & STx3.1        & 1              & 1                 & 10$^{4}$\\
   12     & 3.3           & 3.3                      & 3.3                      & STx3.3        & 1              & 1                 & 10$^{4}$\\
   13     & 2.5           & 2.5                      & 2.5                      & STx2.5        & 1              & 1                 & 10$^{5}$\\
   14     & 2.5           & 2.5                      & 2.5                      & STx2.5        & 1              & 1                 & 10$^{6}$\\
   15     & 2.5           & 2.5                      & 2.5                      & STx2.5        & 10             & 1                 & 10$^{4}$\\
   16     & 2.5           & 2.5                      & 2.5                      & STx2.5        & 100            & 1                 & 10$^{4}$\\
   17     & 2.5           & 2.5                      & 2.5                      & STx2.5        & 1              & 10                & 10$^{4}$\\
   18     & 2.5           & 2.5                      & 2.5                      & STx2.5        & 1              & 100               & 10$^{4}$\\
   19     & 2.5           & 2.5                      & 2.5                      & SW98          & 1              & 1                 & 10$^{4}$\\
   20     & 2.5           & 2.5                      & 2.5                      & WE04          & 1              & 1                 & 10$^{4}$\\
   \hline
    \end{tabular}
    \end{center}
\end{table*}

\section{Sensitivity of chemistry to variations of the
UCL\_PDR model parameters}\label{sec:resu}

In this section, we present the trends the chemistry is showing
with respect to the changes of the parameters described in Section
\ref{sec:param}. We have selected about 25 molecules for closer
study, either for their likely detectability or for their chemical
interest. We have arbitrarily fixed the limit of detectability of
a molecule X to be with a relative abundance of
n(X)/n$_{H}=1\times 10^{-12}$ (where n$_{H}=$n(H) $+$
2n(H$_{2}$)), a criterion roughly satisfied in our own galaxy. The
rest of our paper thus deals specifically with species H$_{2}$, H,
electron, C$^{+}$, C, CO, CS, SO, SO$_{2}$, OCS, H$_{2}$CS,
H$_{2}$S, C$_{2}$H, OH, H$_{2}$O, H$_{3}$O$^{+}$, HCO$^{+}$, CH,
NH$_{3}$, CN, HCN, HNC, H$_{2}$CO, H$_{2}$CN and CH$_{3}$OH. We
additionally exclude methanol from this study because this species
is mainly form on grains and such reactions are not included in
the model we have used. There is only one (minor) gas-phase
formation route that we include in our models but since we are not
properly modelling its formation, we do not think we can say
anything in this study about the variation of methanol abundances
with parameters such metallicity.

\subsection{Gas number density, FUV radiation field and cosmic ray ionisation rate}\label{subsec:dens}

As recently published by \citet{Krip10, Croc11}, HCN and HCO$^{+}$
lines have been detected in high-metallicity environments such as
ETGs. These detections show the likely presence of gas denser than
that traced by CO since to excite the transitions HCO$^{+}$(J=1-0)
and HCN(J=1-0), part of the gas density needs to be close to the
corresponding line critical densities of 3.4$\times 10^{4}$
cm$^{-3}$ and 2.3$\times 10^{5}$ cm$^{-3}$, respectively. We
investigated thus the change in high-metallicity chemistry when
the density increased by a factor of 10 (Model 13 in Table
\ref{tab:2}) or 100 (Model 14) as compared to the selected
standard value of $10^{4}$ cm$^{-3}$ (Models 0). Not surprisingly,
changes in gas density affect high-metallicity chemistry in the
same way as found for solar-metallicity chemistry. Detailed
analysis of such changes can be found in \citet{Baye09a}.
Similarly, the influence of the FUV radiation field and the cosmic
ray ionisation rate on high-metallicity chemistry is the same as
for solar-metallicity chemistry. Since results have been already
published, for the FUV radiation field influence we refer the
reader to Figure 5 and Table 8 from \citet{Baye09a}. For a
detailed influence of the cosmic ray ionisation rate on the
chemistry, we refer the reader to e.g. \citet{Papa10a, Baye11b,
Meij11}.

\subsection{Metallicity}\label{subsec:meta}

\begin{figure*}
    \centering\includegraphics[width=14cm]{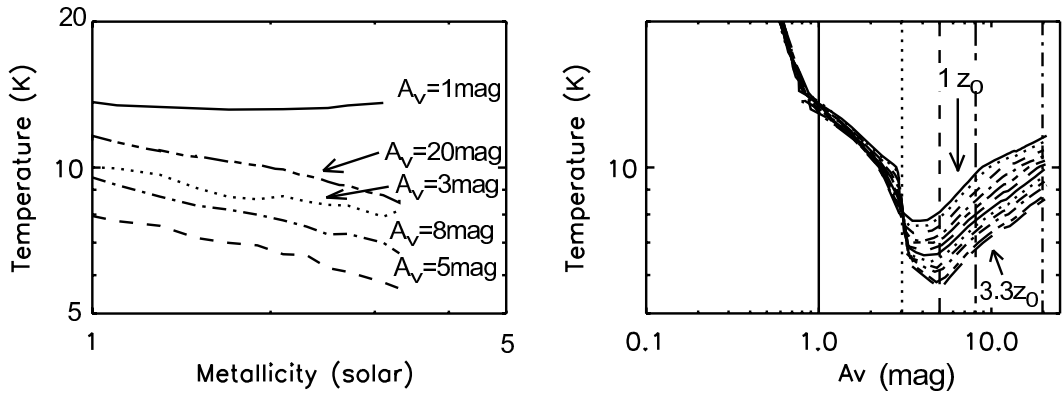}
    \caption{Temperature profiles with respect to \emph{Left:} the
    metallicity (in z$_{\odot}$) and \emph{Right:} the optical depth
    A$_{\rm v}$ (in mag). On the right hand side plot, vertical black
    lines refer to the 5 A$_{\rm v}$ selected and seen on the left hand
    side plot (see Section \ref{sec:param}). We kept
    the same line style in both plots as well as in Fig. \ref{fig:2}
    i.e. A$_{\rm v}$ of 1 mag is represented by solid lines, A$_{\rm v}$
    of 3 mag is shown by dotted lines, A$_{\rm v}$ of 5 mag is designated
    by dashed lines, A$_{\rm v}$ of 8 mag is represented by
    dashed-dotted lines and A$_{\rm v}$ of 20 mag is described by black dotted-dotted-dashed lines. The
    curves seen on the right hand side plot are obtained for various
    metallicity (from solar to 3.3 solar by steps of 0.2), using a different
    line style for each metallicity value from 1 solar (solid lines) to 3.3.
    solar (large dashed lines).}\label{fig:1}
\end{figure*}

We have plotted in Fig. \ref{fig:1} the temperature distribution
obtained from the self-consistent thermal balance, as a function
of the metallicity (left hand side) for the five A$_{\rm v}$
selected previously. We have also plotted the temperature
distribution as a function of A$_{\rm v}$ for the range of studied
metallicities (i.e. from solar to 3.3 z$_{\odot}$ by step of 0.2
z$_{\odot}$, see right hand side of Fig. \ref{fig:1}). The curves
on the left hand side of Fig. \ref{fig:1} show the same general
behaviour for A$_{\rm v}>$ 1 mag i.e. decreasing with increasing
metallicity, reaching a minimum of about 6 K for A$_{\rm v}=$ 5
mag and going up again for higher extinctions. This increase of
temperature after A$_{\rm v}=$ 5 mag, already mentioned in
\citet{Baye11b}, is due to the fact that at high visual
extinctions the C and CO lines become optically thick and hence
are less able to cool the gas whilst the other sources of heating
(such as the cosmic rays) which dominate the heating at these
depths (60\% of the total heating) remains constant. At A$_{\rm
v}=$ 5 mag, the cooling is actually dominated by CS (35.7\%),
followed by H$_{2}$O (20\%) and $^{13}$CO (13\%) which show again
the importance of the newly added radiative cooling mechanisms
(see Section \ref{sec:mod} and \citealt{Baye11b}). The temperature
curve at A$_{\rm v}=$ 1 mag appears insensitive to metallicity
changes; increasing the metal content of the ISM thus does not
seem to have significant impact on the total heating or cooling
mechanisms of the UV irradiated part of the ISM. This is indeed
expected since increased metallicity enhances similarly the PAH
and dust photo-electric heating as well as the CO cooling. The
ratio between heating and cooling processes stays thus roughly the
same from a metallicity value to another. No additional radiative
contribution to the total cooling from molecules are seen at
A$_{\rm v}=1$ mag since at such depth, molecules are mainly
dissociated by the FUV radiation field. Generally, the
temperatures seen in Fig. \ref{fig:1} stay below 20 K whatever the
metallicity and the A$_{\rm v}$, ensuring a rich chemistry.

\begin{figure*}
    \centering
    \includegraphics[width=14cm]{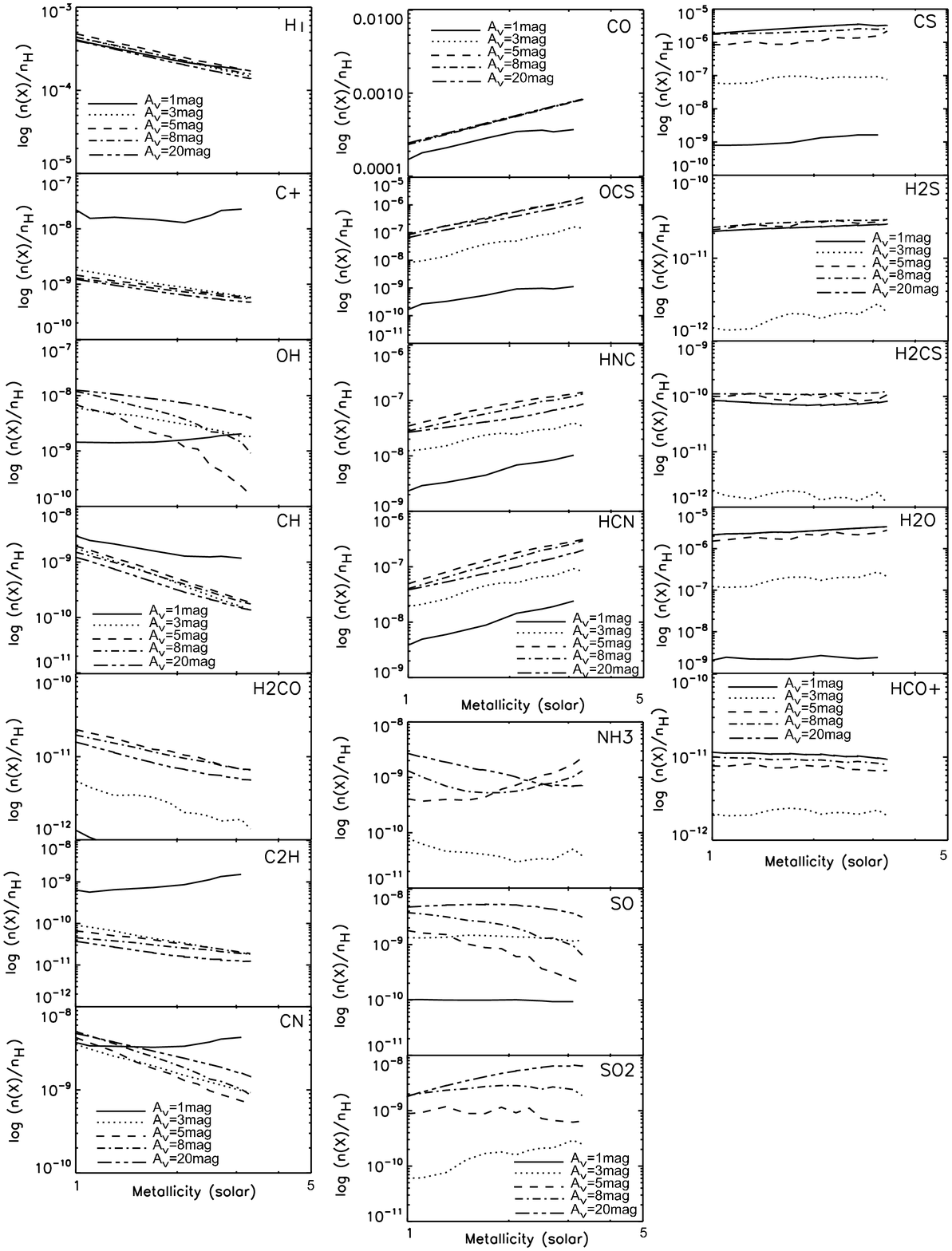}
    \caption{Fractional abundances (n(X)/n$_{\rm H}$)
    of species X the most sensitive to the changes in metallicity,
    in logarithmic scale, where n$_{\rm H}$ is the total number of
    hydrogen atoms, with respect to the metallicity (in z$_{\odot}$,
    see text in Subsect. \ref{subsec:meta}). These plots summarize
    the chemistry changes seen for Models 0 to 12 with A$_{\rm v}$ of 1 mag,
    3, 5, 8 and 20 mags (see caption of Fig. \ref{fig:1}). The lines for
    A$_{\rm v}$ of 1 mag in the NH$_{3}$ and the SO$_{2}$ do not appear because
    they are well below our limit of detectability. On the left and right (top)
    hand sides of this figure, the species fractional abundances
    decrease and increase, respectively, with the metallicity
    changes. On the bottom right hand side are the outsidders
    species behaving differently from an A$_{\rm v}$ to another in
    response to the metallicity changes.}\label{fig:2}
\end{figure*}

Increasing the metallicity (see Fig. \ref{fig:2}) has a different
impact from one species to another. When increasing the
metallicity, we can disentangle three different behaviours in the
chemical abundances. The species seen on the left hand side of
Fig. \ref{fig:2} decrease by a factor of 3 to 10 with increasing
metallicity whereas those plotted on the upper right hand side of
Fig. \ref{fig:2} increase by a factor of 3 to 15 with increasing
metallicity. H$_{2}$, C, electron, CS, H$_{2}$S, H$_{2}$CS,
H$_{2}$O, H$_{3}$O$^{+}$, HCO$^{+}$ and H$_{2}$CN do not show
significant variations and thus can be considered as species
insensitive to metallicity increase from the solar value.
Interestingly, one can also add to this list of `constant'
fractional abundances those of OH and CN for the case A$_{\rm
v}=1$ mag only (see black solid lines in Fig.\ref{fig:2}). Outside
these three categories is SO with an increased fractional
abundance at A$_{\rm v}=1$, 3 and 20 mag but otherwise decreasing;
SO$_{2}$ with a decreased fractional abundance at A$_{\rm v}=5$
and 8 mag but otherwise increasing; NH$_{3}$ with a decreasing
fractional abundances with increasing metallicity for A$_{\rm
v}=3$ and 20 mag but otherwise increasing for A$_{\rm v}=5$ mag
and increasing then decreasing for A$_{\rm v}=8$ mag (see bottom
right hand side of Fig. \ref{fig:2}). C$^{+}$ and C$_{2}$H at
A$_{\rm v}=1$ mag have smaller variations than those seen for SO,
SO$_{2}$ and NH$_{3}$ but the fact that the C$^{+}$ and C$_{2}$H
fractional abundances are high (i.e. above 10$^{-10}$) makes these
more moderate variations as meaningful as those seen for SO,
SO$_{2}$ and NH$_{3}$.

Complementary results on the variation of the fractional
abundances of chemical species with changes in metallicity can be
found in \citet{Baye09a}, where we showed that, CS in particular,
is sensitive to changes in metallicity. This apparent
contradiction with our current results can be explained by the
fact that the models in \citet{Baye09a} where for low metallicity
whereas in the present study we investigate super-solar
metallicities appropriate for ETGs. In addition, in
\citet{Baye09a} the FUV radiation field is set three times higher
than in our present study, which has a dramatic effects on the CS
fractional abundance (see their Table 8).

It is on average very difficult to determine the reason for those
changes in the chemistry caused by the metallicity because the
effects are non-linear involving changes in both total elements
and dust-to-gas ratio. However, the first category of species can
be understood as actually a result of what happen to the species
of the second category when the metallicity increases. For
instance, species such as HCN and HNC (category 2) form
predominantly from CN, a species which decreases its abundance at
high metallicity (first category). This decrease is linked to the
decrease of H at high metallicity, H being involved in one of the
main formation routes of CN. The decrease of the CN abundance is
also reinforced by the destruction route CN $+$ H$_{2}$ where
H$_{2}$ remains high whatever the metallicity. Thus, if HCN and
HNC abundances increase with metallicity, this must cause the CN
abundance to drop, and vice-versa. Similar chemical routes exist
between CO, C$_{2}$H and H$_{2}$CO. Detailed chemical analysis
following the formation and destruction routes of some of the
species included in our models has been performed and confirms the
present interpretation of these categories. The behaviour of
species whose abundances are independent of the change in
metallicity can be explained by the fact that those species are
both formed and destroyed by exchange reactions that equally
depend on metallicity and the net result is that no change occurs
due to metallicity changes. For instance, H$_{2}$S is destroyed
mainly by oxygen which increases with the metallicity, but is
formed mainly by electrons recombining with H$_{3}$S$^{+}$,
H$_{2}$S$^{+}$ and H$_{2}$S$_{2}^{+}$, reactions which also
increase with metallicity. This results in a simultaneous increase
in both formation and destruction, which leaves the abundance of
H$_{2}$S unchanged. More generally, an increase in metallicity
leads to a decrease in H, C$^{+}$ and molecular ions via
dissociative recombination, and an increase in electrons. One
notes also that, as a secondary effect of the increase in
metallicity, the secondary photon process is enhanced due to an
increase of total hydrogen nuclei.

\subsection{$\alpha-$element enhancement}\label{subsec:alpha}

\begin{figure*}
    \centering
    \includegraphics[width=10cm]{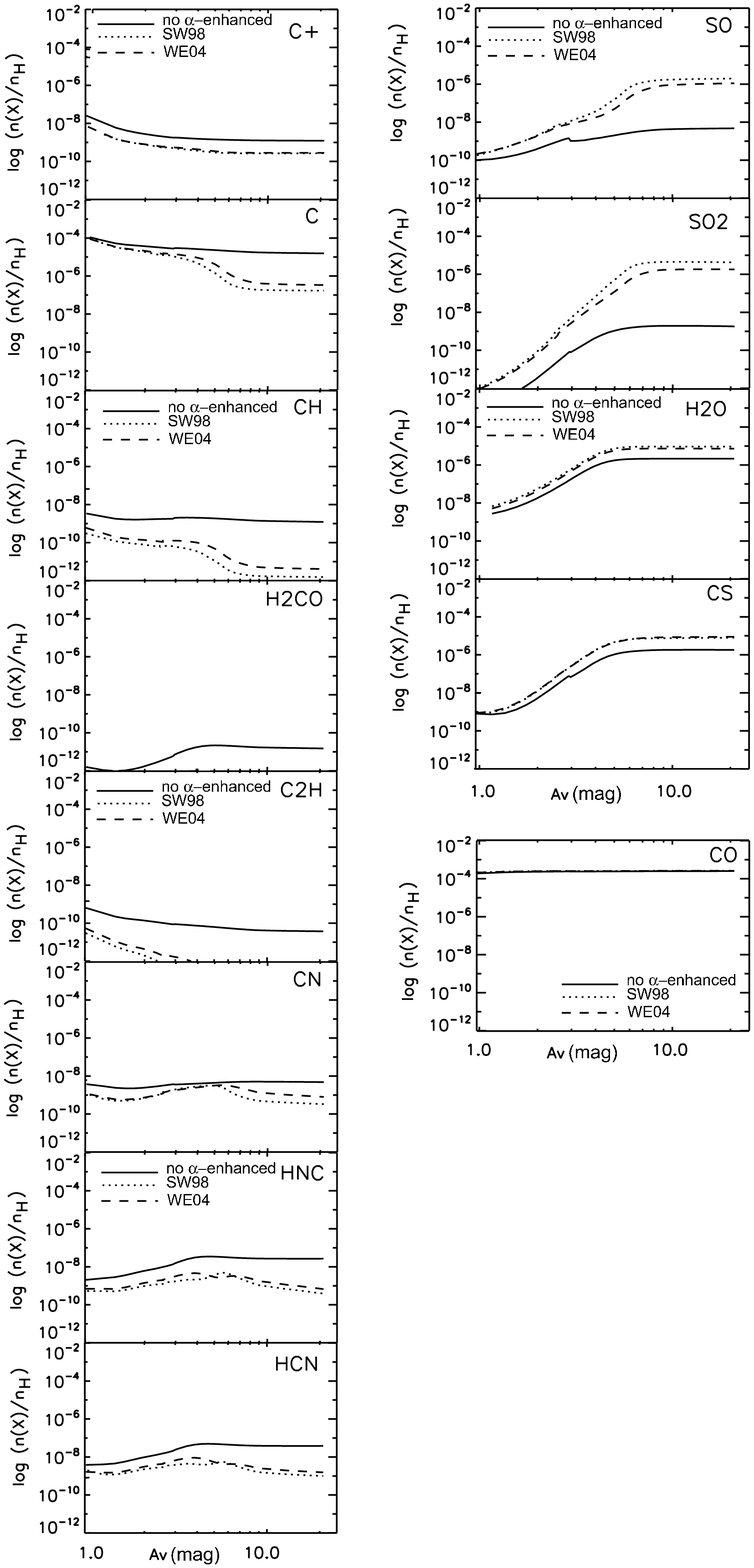}
    \caption{Fractional abundances (n(X)/n$_{\rm H}$)
    of species X the most sensitive to the changes in
    $\alpha-$element enhancement (see caption of Fig.
    \ref{fig:2}). The plain black lines correspond to
    results from Model 8 at 2.5 metallicity and without
    $\alpha-$element enhancement, whereas black dotted
     and black dashed lines represent Model 19 and 20, respectively
      both having a metallicity of 2.5 z$_{\odot}$ but different scenario of
    $\alpha-$element enhancement (see text in subsection
    \ref{subsec:alpha} and Table \ref{tab:3}).}\label{fig:3}
\end{figure*}

This is the first time that high-metallicity chemistry has been
studied with respect to the changes in $\alpha-$elements content
(see Fig. \ref{fig:3}). As described in Section \ref{sec:param}
and seen in Table \ref{tab:2}, we have used in the present work
two mixtures of $\alpha-$elements, one corresponding to the
approach of \citet{Sala98} with different factors for each
$\alpha-$element (Models 19 in Table \ref{tab:3}) and one from
\citet{Weis04} who increase by a factor of 0.4 dex all the
$\alpha-$elements (Models 20 in Table \ref{tab:3}).

We can divide molecular species in three categories: those
insensitive to the changes in $\alpha-$elements i.e. whose
fractional abundances do not change by more than a factor of 2:
CO, H, OH, NH$_{3}$, H$_{2}$S, OCS, H$_{3}$O$^{+}$, HCO$^{+}$; the
species which have their fractional abundances increasing by a
factor between 3 and more than 3 orders of magnitude when the
$\alpha-$elements are enhanced (see right hand side of Fig.
\ref{fig:3}) i.e. SO, SO$_{2}$, CS and H$_{2}$O. Finally, the
third category includes CN, HCN, HNC, CH, C$_{2}$H, H$_{2}$CS,
C$^{+}$, C and H$_{2}$CO. These species decrease in fractional
abundances with the $\alpha-$element enhancement, by up to 5
orders of magnitude for C$_{2}$H (see left hand side of Fig.
\ref{fig:3}). We have not shown in Fig. \ref{fig:3} the evolution
of the fractional abundance of H$_{2}$CN with respect to the
change in $\alpha-$elements because its fractional abundance drops
close or below our limit of detectability. Hence this species,
even if sensitive to the changes in $\alpha-$elements is not
likely to be detectable. However a stringent non-detection could
be indicative. Similarly, for C$_{2}$H, CH at A$_{\rm v}>$ 8 mag,
and H$_{2}$CO, the decrease of fractional abundance is so
pronounced that they fall under the assumed limit of
detectability. Otherwise, the response of the fractional abundance
of the species belonging to the second and third categories make
them quite sensitive, hence potential good tracers of
$\alpha-$element enhancement. For those species whose fractional
abundance is above 10$^{-10}$ whatever the optical depth, they are
likely detectable.

Another interesting aspect is the potential difference in the
chemistry between the two scenarios of $\alpha-$element
enhancement. The species showing the most contrast in their
fractional abundances are C$_{2}$H, H$_{2}$S, CH and C (increase
of a factor greater than 2 between Models 19 and 20) and SO and
SO$_{2}$ (decrease by a factor greater than 2). Here, only species
showing a fractional abundance varying by a factor of more than
five between the scenarios of $\alpha-$element enhancement (i.e.
only C$_{2}$H) may be useful for distinguishing the \citet{Sala98}
scenario (i.e. various factors applied to the $\alpha-$elements)
from the \citet{Weis04} one (i.e. same factor of +0.4 dex applied
to all $\alpha-$elements). However the fractional abundance of
C$_{2}$H in Models 19 and 20 is very low, whatever the optical
depth which makes this species unlikely to be detectable. To
summarize, it appears thus not possible to use models to
disentangle the two scenarios of $\alpha-$element enhancement.

\section{Molecular tracers of
high-metallicity environments}\label{sec:ana}

\subsection{Fractional abundances}

In this section, we attempt to identify the best chemical tracers
of high-metallicity environments and to provide corresponding line
brightnesses (in erg s$^{-1}$ cm$^{-2}$ sr$^{-1}$) useful for
future observational programmes. We restrict the number of species
we analyse to those (i) belonging to the list of the most
sensitive species seen in Subsection \ref{subsec:meta} and
\ref{subsec:alpha} i.e. to those having a fractional abundance
changing by more than a factor of 5 with changes in metallicity or
in $\alpha-$elements; to those (ii) having a fractional abundance
at least 10 times higher than our arbitrarily fixed limit of
detectability; and to those (iii) having at least one transition
previously observed in extragalactic environments. This leads to
the detailed study of C$^{+}$, C, CO, C$_{2}$H, CN, HCN, HNC, OCS,
CS, SO, SO$_{2}$ and H$_{2}$O.

From Subsection \ref{subsec:meta}, species likely to be the best
at probing changes in metallicity are C$^{+}$, C$_{2}$H, CN, HCN,
HNC, OCS, CO. Similarly, the most likely chemical tracers for
determining whether there is an enhancement in $\alpha-$elements
are C$^{+}$, C, CN, HCN, HNC, SO, SO$_{2}$, H$_{2}$O and CS.
Potentially, a non-detection of C$_{2}$H may also be used to
indicate the presence of an enhanced $\alpha-$elements
environment.

In Table \ref{tab:4}, we provide values of fractional abundance
ratios involving the above species with CS and CO for Models 0,8
and 12 at 5 different extinctions. The fractional abundance of CS
and CO have been selected as denominators for the ratios because
CS is shown to be rather insensitive to changes in metallicity and
CO insensitive to the presence of an $\alpha-$element enhancement
(see Section \ref{sec:resu}).

There is no strong variation seen in the ratios for A$_{\rm v}\leq
3$ mag. Therefore, this range of extinctions does not have an
observational interest. The ratios showing the largest variations
when the metallicity increases are C$^{+}$/CO, SO/CO, CN/CO and
CN/CS, with a factor of differences between Models 0, 8 and 12, at
A$_{\rm v}=5$ mag, of 160, 36, 22 and 16, respectively. These
ratios thus might be considered as the best and most likely
observable probes of the metallicity changes in metal-rich
environments. In addition, the ratios showing the largest
variations when there is an $\alpha-$element enhancement are
C$_{2}$H/CS, SO$_{2}$/CO, SO/CO, HCN/CS and HNC/CS with factors of
difference between Models 8 and 19, at A$_{\rm v}= 5$ mag, of
$>$5000, 1049, 669, 117 and 71, respectively. At higher A$_{\rm
v}$, C/CO becomes also a good indicator as it differs between
Models 8 and 19 by more than a factor of 55. These ratios might
thus be considered as the most reliable probes of an
$\alpha-$element enhancement in metal-rich environments.

\subsection{Line brightness ratios}

The LVG treatment used in the UCL\_PDR code to calculate the line
brightness requires collisional rates for the studied species
(typically with H$_{2}$). Unfortunately, for C$_{2}$H, these
collisional rates are not available. We therefore do not have line
brightness predicted by our models for this species. We also do
not calculate line brightness for SO$_{2}$ and H$_{2}$O because
they contain a very large number of level populations which are
required for a proper treatment, and, as a consequence, an
unreasonable amount of computational time.

In Fig. \ref{fig:4}, for Models 0 and 12, we have therefore
plotted line brightness ratios for C$^{+}$/CS, CN/CS, HCN/CS,
HNC/CS, CO/CS and OCS/CS as a function of A$_{\rm v}$. We
restricted our plots to line ratios involving transitions
observable simultaneously using large backends on telescopes such
as IRAM-30m or ALMA, i.e. to those with close frequencies such as
HCN(1-0)/CS(2-1), HNC(3-2)/CS(5-4), OCS(28-27)/CS(7-6), etc. In
the right hand side of Fig. \ref{fig:4}, we see the distribution
of the line brightness ratios with respect to A$_{\rm v}$ for
Model 0 only. On the left hand side of Fig. \ref{fig:4}, we see
the comparison between Models 0 (black lines) and 12 (grey lines)
i.e. the comparison between solar and 3.3 z$_{\odot}$ metallicity,
for only a selected number of line brightness ratios. In Fig.
\ref{fig:5}, we have plotted the C$^{+}$/CO, C/CO, CN/CO, HCN/CO,
HNC/CO and CS/CO ratios. Similarly to figure \ref{fig:4}, on the
right hand side of Fig. \ref{fig:5}, the results from Model 8 are
displayed whereas on the left hand side of Fig. \ref{fig:5}, we
see the comparison between Model 8 (black lines, without
$\alpha-$element enhancement) and Model 19 (grey lines, with
$\alpha-$element enhancement). The top panels in both figures
present the ratios between well-known ISM cooling lines i.e.
C$^{+}$, OI (146 $\mu$m), OI (63 $\mu$m) , OI (44 $\mu$m), CI (610
$\mu$m), CI (370 $\mu$m) and CI(230 $\mu$m) and the CS(2-1) line
(Fig \ref{fig:4}) or the CO(1-0) transition (Fig \ref{fig:5}). The
CS(2-1) and CO(1-0) transitions were selected as denominators
firstly because the CS and CO fractional abundances are
insensitive to changes in metallicity and presence of
$\alpha-$elements, respectively, and secondly, because these two
lines have easily detectable emission at 3mm in external galaxies
(see e.g. \citealt{Isra01, Isra03, Baye04, Baye06, Baye09c,
Alad11}). One notes that OI is not amongst the species respecting
conditions i), ii) and iii), but we have nevertheless plotted its
OI (146 $\mu$m), OI (63 $\mu$m) and OI (44 $\mu$m) line
brightnesses because they are routinely observed in extragalactic
environments. Additional line ratios will soon be available online
(see http://www.homepages.ucl.ac.uk/~ucapdwi/interface).

The results derived from figures \ref{fig:4} and \ref{fig:5} can
be summarized as follow:
\begin{itemize}
\item The increase in metallicity impacts similarly for all the
ratios involving the well-known ISM cooling lines with respect to
the CS(2-1) line, i.e. these ratios have higher values up to a
factor of 20 (for OI (44 $\mu$m)/CS(2-1)) as compared to their
value in the case of solar metallicity (i.e. Model 0). Similarly,
the OCS/CS line brightness ratios increase with the increase of
metallicity, whatever the transition considered. For well-known
ISM cooling line ratios, the increase between solar and 3.3
z$_{\odot}$ is constant with respect to the A$_{\rm v}$ whereas
for the OCS/CS ratio, this increase decreases when A$_{\rm v}$
increases. This allow us to propose this ratio as powerful
diagnostics of either the metallicity or the opacity. Indeed, if
the OCS(8-7)/CS(2-1) line brightness ratio is measured in external
galaxies to be for instance 0.1, this corresponds either to a gas
metallicity of 3.3 solar and a rather translucent gas (with an
A$_{\rm v}\sim 1.8$ mag) or the metallicity of the gas is solar
and the A$_{\rm v}\sim$ 4 mag. To break this metallicity-opacity
degeneracy, independent estimates of the A$_{\rm v}$ are required.
To do so, molecular isotopologues detections are essential.
Possible way to break this degeneracy is to use a set of line
ratios, providing they involve lines with similar critical
densities, allowing us to assume they originate from the same gas
component. For example, a combination of H$_{2}$CS(7-6),
HCO$^{+}$(3-2), H$_{2}$CO(3-2), SO(5-4) and C$_{2}$H(3-2) compared
with CS(5-4) would be useful since those lines, close in frequency
to the CS(5-4) transition, have also similar critical densities
(of about $3-8 \times 10^{5}$ cm$^{-3}$). For the other ratios
seen in Fig. \ref{fig:4} (i.e. CN/CS, HCN/CS, HNC/CS and CO/CS),
but excluding OCS(8-7)/CS(2-1) the metallicity boost to 3.3
z$_{\odot}$ increases the ratios as compared to the solar case for
A$_{\rm v}\leqslant 3-4$ mag up to a factor of 8 for
HCN(1-0)/CS(2-1) and HCN(4-3)/CS(7-6). Then, at A$_{\rm v}\sim 5$
mag (A$_{\rm v}\sim 2.8$ mag for the CN\_1/CS(2-1)), the two
metallicity models give similar ratios. At these extinctions, it
will thus not be possible to use the models predictions for
estimating the gas metallicity. Finally, at A$_{\rm v}\geqslant 5$
mag, the ratios for the solar and 3.3 z$_{\odot}$ metallicities
are decoupled again, with lower ratio values for the 3.3
z$_{\odot}$ metallicity case than for the solar one. At high
extinctions, however the difference between the two modelled
ratios might not be large enough to be detectable, leading to a
rather uncertain estimate of the metallicity. Here, we have
considered observational uncertainties (calibration, pointing,
etc) of 20\%, i.e. to be detectable, we need a difference between
ratios greater than a factor of 5. Two exceptions stand in the
HCN/CS and HNC/CS ratios involving transitions at about 300 GHz
frequency: if the A$_{\rm v}$ is estimated independently to be of
about 8 mag, then the metallicity can be determined (factor of 8
difference between the ratios at solar and 3.3 solar metallicity).
One notes that line brightness CO/CS ratios are \emph{not} the
best ratios to use for estimating the metallicity since the
difference between the two model predictions are not as large as
those seen for the HNC/CS and HCN/CS ratios (the latter ones being
a much better indicator of metallicity). \item In high-metallicity
environments such as ETGs (here we assume the metallicity to be at
2.5 z$_{\odot}$ - see Fig. \ref{fig:5}), it may possible to
confirm the presence of $\alpha-$element enhancements by measuring
line brightness SO/CO ratios. Indeed these ratios show the
strongest increase of values between Models 8 and 19, especially
for A$_{\rm v}\geqslant 4$ mag and this, whatever the transition
studied. An $\alpha-$element enhancement decreases the rest of the
line brightness ratios by a factor 0.5 to 5. One notes that the
decrease of the ratios in the case of low frequency ratios such as
HCN(1-0)/CO(1-0), HNC(1-0)/CO(1-0) and CN\_1/CO(1-0) is not large
enough for ensuring a confirmation of $\alpha-$element enhancement
whereas for the ratio involving higher frequency line brightnesses
such as HCN(4-3)CO(3-2), HCN(4-3)/CO(3-2), especially for A$_{\rm
v}>5$ mag, this is becoming possible.

\end{itemize}

In Figure \ref{fig:6}, we plot theoretical and observed CO line
ratios for ETGs. CO(1-0), CO(2-1), $^{13}$CO(1-0), $^{13}$CO(2-1),
HCO$^{+}$(1-0) and HCN(1-0) data have been published for a limited
number of gas-rich ETGs \citep{Krip10, Youn11}. From the
$^{12}$CO/$^{13}$CO ratio and assuming a Milky Way value of the
abundance ratio $^{12}$C/$^{13}$C=70 (see \citealt{Mart10} and
references therein), using the Eq. 8 in \citet{Baye04}, we have
obtained an A$_{\rm v}$ for the $^{12}$CO gas of 5.5 mag, 8.0 mag,
22 mag and 23.7 mag for NGC4150, NGC3032, NGC4526 and NGC4459,
respectively from $^{12}$CO(1-0)$/^{13}$CO(1-0) ratios of 13.67,
9.46, 3.46 and 3.20, respectively from \citealt{Krip10} and
\citealt{Youn11}.

On the left hand side of Fig. \ref{fig:6}, we compare the CO
beam-dilution corrected observations to our model predictions. In
this plot, we have plotted the variations of the CO(2-1)/CO(1-0)
line brightness ratio with respect to A$_{\rm v}$ for a solar
(black line) and a 3.3z$_{\odot}$ metallicity (grey lines).
Crosses represent the current observed CO(2-1)/CO(1-0) line
brightness ratios for four observed gas-rich ETGs (derived from
the beam corrected velocity-integrated line intensity ratios
displayed in \citealt{Krip10} data. For \citealt{Youn11} data, we
used beam dilution correction method described in
\citealt{Croc11}). We can see that the variation of the
CO(2-1)/CO(1-0) ratio with the increase of metallicity is not
significant enough to reliably constrain the metallicity of these
gas-rich ETGs, even if the A$_{\rm v}$ has been determined using
$^{13}$CO(1-0) data. However, one sees that the line brightness
CO(6-5)/CO(1-0) ratio can be useful as a CO line ratio alternative
to estimate metallicity. The typical error on each observed
CO(2-1)/CO(1-0) ratio is of $\sim$ 12\% (see \citealt{Krip10,
Youn11}). If similar error can be obtained on the line brightness
CO(6-5)/CO(1-0) ratio, then, from Fig \ref{fig:6}, reliable
metallicity estimate may be possible to obtain, especially if the
gas is strongly opaque as in NGC4526 and NGC4459 (whose Av is
outside the plot). CO(J--J-1)/CO(1-0) lines at high frequency seem
thus a better indicator of the metallicity in high metallicity
environments than lower frequency CO line ratios, but, in any
case, ratios such as HCN/CS and HNC/CS are more powerful
diagnostics. Future work \citep{Davi12} will explore this issue
with new and recently released observational data \citep{Croc11}.

\section{Conclusions}\label{sec:con}

\begin{figure*}
    \centering
    \includegraphics[width=11cm]{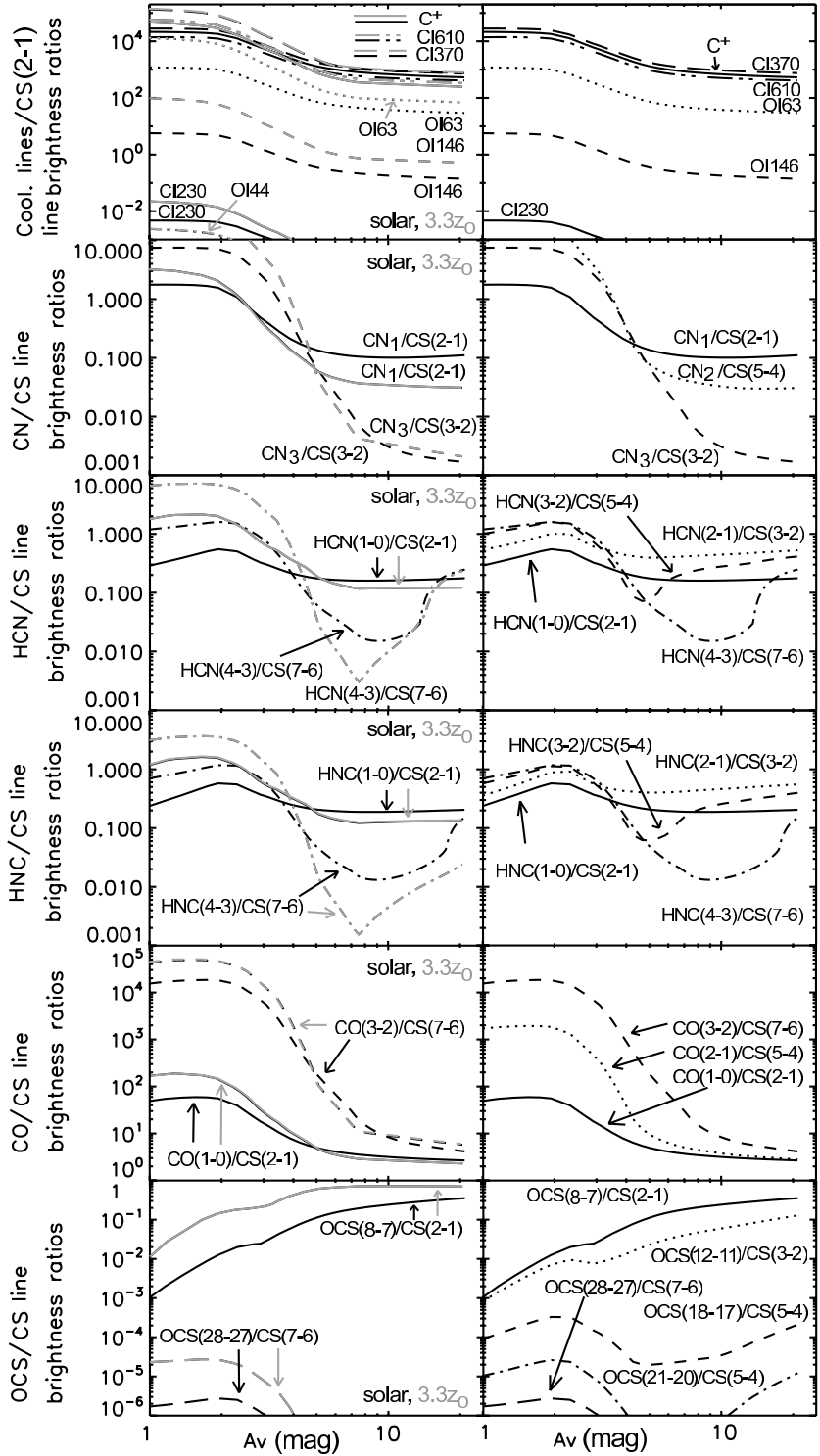}
    \caption{Line brightness ratios for various species with respect to the CS lines (study of
    the metallicity). We present plots from top to bottom of the far-infrared fine structure
    cooling lines : C$^{+}$/CS(2-1), OI(44$\mu$m)/CS(2-1), OI(63$\mu$m)/CS(2-1),
    OI(146$\mu$m)/CS(2-1), CI(230$\mu$m)/CS(2-1), CI(370$\mu$m)/CS(2-1), CI(610$\mu$m)/CS(2-1),
    CN/CS, HCN/CS, HNC/CS, CO/CS and OCS/CS ratios. The CN$_{1}$, CN$_{2}$ and CN$_{3}$
    transitions represent emission at 113.490 GHz, 226.874 GHz and 340.247 GHz,
    respectively. Every line involved in a ratio is mentioned on the right
    and left hand side plots.
    \emph{Left:} Comparison between Models 0 (black lines) and Model 12 (grey lines).
    \emph{Right:} Predictions for Model 0. The ratio plotted on the left hand side are
    the ones plotted, in the same line style, on the right hand side.}\label{fig:4}
\end{figure*}

\begin{figure*}
    \centering
    \includegraphics[width=11cm]{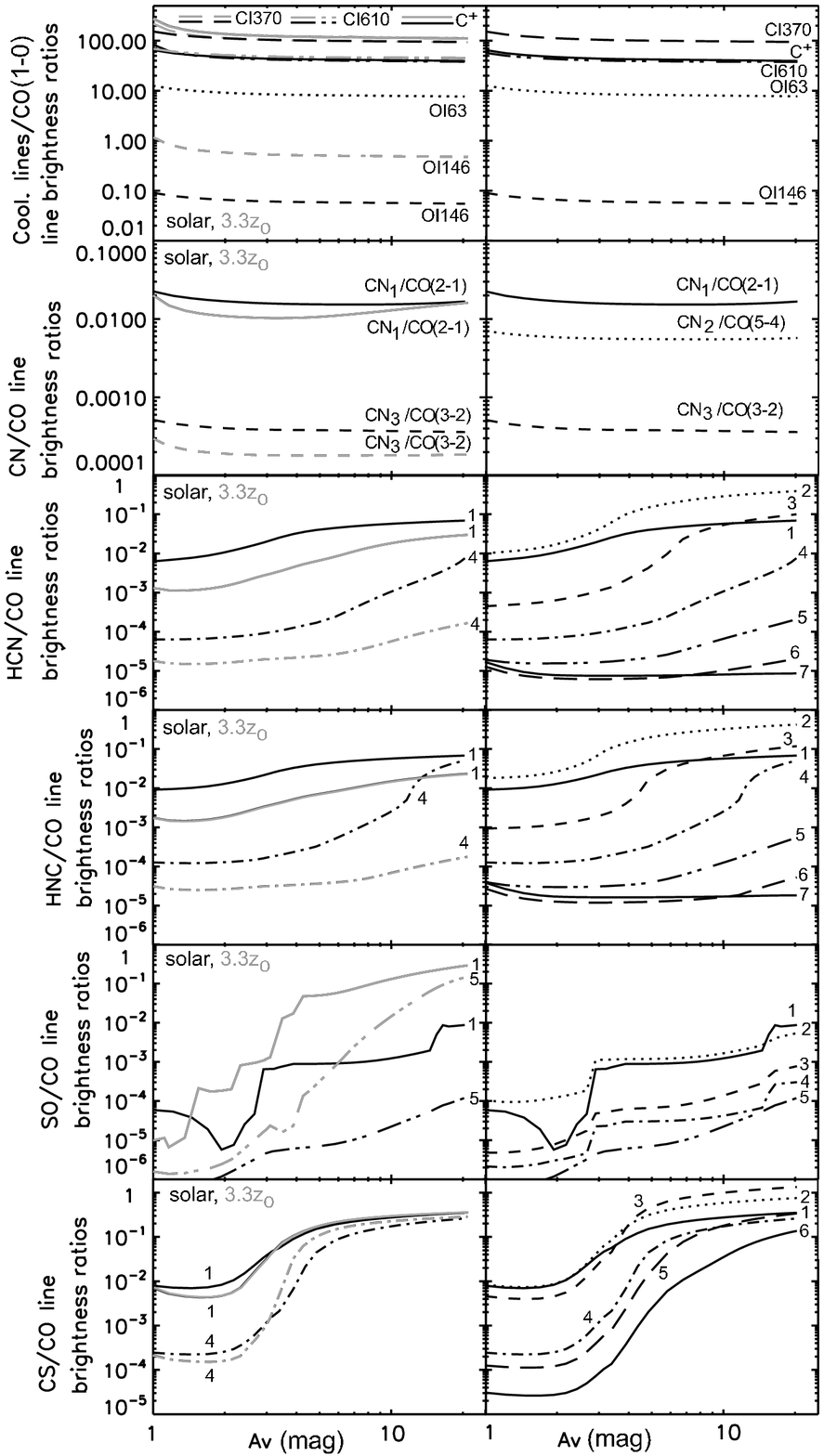}
    \caption{Line brightness ratios for various species with respect to the CO lines (study
    of the $\alpha-$element enhancement). We present plots from top to bottom of the far-infrared
    fine structure cooling lines : C$^{+}$/CO(1-0), OI(63$\mu$m)/CO(1-0),
    OI(146$\mu$m)/CO(1-0), CI(370$\mu$m)/CO(1-0), CI(610$\mu$m)/CO(1-0), CN/CO,
    HCN/CO, HNC/CO, SO/CO and CS/CO ratios. The CN$_{1}$, CN$_{2}$ and CN$_{3}$ transitions
    correspond to the CN emission at 113.490 GHz, 226.874 GHz and 340.247 GHz, respectively.
    The numbers 1 to 7 in the plot of the HCN/CS line brightness ratios correspond respectively
    to HCN(1-0)/CO(1-0), HCN(2-1)/CO(1-0), HCN(3-2)/CO(2-1), HCN(4-3)/CO(3-2), HCN(5-4)/CO(4-3),
    HCN(6-5)/CO(5-4) and HCN(7-6)/CO(6-5). Similar lines ratios are plotted for
    HNC/CS. For the plot of the SO/CO, we have the SO emission at 1=109.252 GHz/CO(1-0),
    2=246.404 GHz/CO(2-1), 3=336.553 GHz/CO(3-2), 4=461.755 GHz/CO(4-3) and 5=609.96 GHz/CO(6-5).
    Finally for the CS/CO plot, we have 1=CS(1-0)/CO(1-0), 2=CS(2-1)/CO(1-0), 3=CS(3-2)/CO(1-0),
    4=CS(4-3)/CO(2-1), 5=CS(5-4)/CO(2-1) and 6=CS(7-6)/CO(3-2). \emph{Left:} Comparison
    between Models 8 (black lines) and 19 (grey lines). \emph{Right:} Predictions from
    Model 8. The ratio plotted on the left hand side are
    the ones plotted, in the same line style, on the right hand side.}\label{fig:5}
\end{figure*}

\begin{figure*}
    \centering
    \includegraphics[width=11cm]{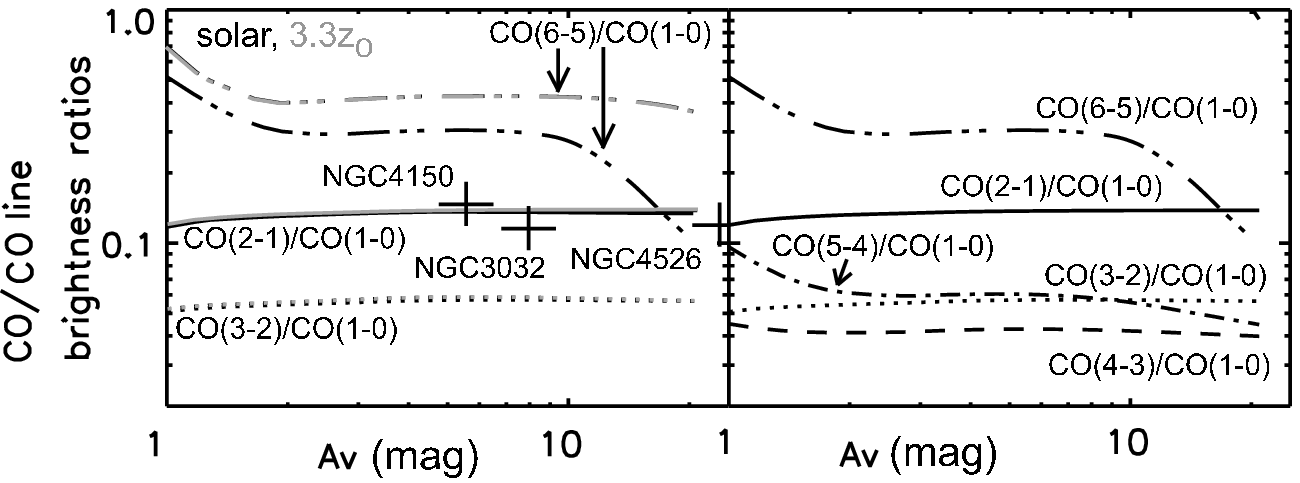}
    \caption{Example of results obtained when comparing observed and predicted CO
    line ratios. \emph{Right:} Results from Model 0 (solar metallicity); \emph{Left:}
    Comparison between Model 0 (black lines) and Model 19 (3.3 z$_{\odot}$ -
    grey lines) predictions for the CO(2-1)/CO(1-0) (solid lines) and the CO(6-5)/CO(1-0)
    (dashed-dotted-dotted lines) line brightness ratios. On top of which we added observational
    values (see black crosses) which represent the data presented in \citealt{Krip10, Youn11},
    taking into account of their errors (crosses show error bars). The A$_{\rm v}$ derived for NGC4459 (23.7 mag) is
    outside the plotted range of optical depths hence does not appear on the figure.}\label{fig:6}
\end{figure*}

The main results of this theoretical study are contained in
Section \ref{sec:resu} which shows how the chemistry evolves with
respect to the changes in various parameters likely to be
appropriate for high-metallicity environments such as ETGs: FUV
radiation field, density, cosmic-ray ionisation rate, metallicity
(increase up to 3 times solar), and an enhancement in the
$\alpha-$elements. The impact of the last two parameters on the
chemistry have never been investigated so far in such detail.
Table \ref{tab:4} and Figs. \ref{fig:2}-\ref{fig:6} are thus of
particular interest, especially when considering the fact that at
high metallicity the influence of the FUV radiation field, density
and cosmic-ray ionisation rate on the chemistry does not differ
significantly from what has been previously shown by e.g.
\citet{Baye09a, Papa10a, Meij11, Baye11b}.

From our study, it appears that the best observable tracers of
metallicity are C$^{+}$, C$_{2}$H, CN, HCN, HNC, OCS and CO
whereas the most likely observable chemical tracers for probing
enhancement in $\alpha-$elements are C$^{+}$, C, CN, HCN, HNC, SO,
SO$_{2}$, H$_{2}$O and CS. Fractional abundances and line
brightness ratios with respect to CS (insensitive to metallicity
changes) and to CO (insensitive to $\alpha-$element changes)
provide more quantitative values, useful for future follow-up
observational programmes. Powerful tracers of metallicity changes
have been identified (from the most sensitive to the least): the
HCN/CS and HNC/CS line brightness ratios at all frequency for a
gas with A$_{\rm v}$ either smaller than 3 mag or of about 8 mag,
line brightness OI/CS(2-1) ratios for a gas with A$_{\rm
v}\leqslant$ 4 mag and OCS/CS line brightness ratios at all
frequencies for gas with A$_{\rm v}\leqslant$ 4 mag. The ideal
line brightness ratio for probing the enhancement of
$\alpha-$elements is the SO/CO whatever the optical extinction and
the frequency of the lines involved. One notes that ratios
involving two CO lines are not the best probes of either
metallicity or $\alpha-$element enhancement. However when
CO(6-5)/CO(1-0) line brightness ratio is used and the error
on the ratio is small, some reasonable estimate of the metallicity
can be obtained. This however may be difficult without a good
treatment of beam dilution effects.

To conclude, if CS data are obtained in sources like ETGs, HCN/CS
line brightness ratios can be calculated and from the results
displayed here, we may be able to estimate the molecular gas
metallicity in these sources which could put strong constraints on
the origin of the gas in ETGs. This work can also be useful in
predicting metallicity independently from previously used (i.e.
mainly optical) methods in other high metallicity environments
such as galaxy centres, etc. Telescopes such as ALMA and the
IRAM-30m telescope are amongst the most privileged instruments to
use for this study since large backends are available, hence
observing simultaneously several molecular lines of CS, HCN, etc
is possible.

\begin{landscape}
\begin{table}
    \caption{Fractional abundance ratios which are likely the best
    tracers of high-metallicity environments for various models (see
    characteristics in Tables \ref{tab:1} to
    \ref{tab:3}).}\label{tab:4}
    \resizebox{24cm}{!}{
    \begin{tabular}{l c c c c c c c c c c c c c c c c}
    ratio&C$^{+}$/CS & C$_{2}$H/CS
    &CN/CS& HCN/CS& HNC/CS& OCS/CS& CO/CS& C$^{+}$/CO&
    C/CO& CN/CO& HCN/CO& HNC/CO&
    SO/CO& SO$_{2}$/CO&
    H$_{2}$O/CO\\
    \hline
    A$_{\rm v}=$ 1 mag &&&&&&&&&&&&&&&&\\
    Model 0 &1.40(1)&4.97(-1)&3.98&5.57&3.44&4.10(-1)&2.68(5)&5.23(-5)&3.70(-1)&1.49(-5)&2.08(-5)&1.29(-5)&5.79(-7)&9.76(-10)&1.41(-5)\\
    Model 8 &2.87&2.10(-1)&1.31&9.69&4.49&1.53&3.50(5)&8.21(-6)&2.50(-1)&3.75(-6)&2.77(-5)&1.28(-5)&2.58(-7)&1.20(-9)&6.47(-6)\\
    Model 12 &3.15&3.00(-1)&1.26&1.11(1)&4.81&1.37&2.85(5)&1.11(-5)&4.20(-1)&4.42(-6)&3.91(-5)&1.69(-5)&1.96(-7)&1.07(-8)&5.17(-6)\\
    Model 19 &2.76&1.00(-2)&6.50(-1)&1.29&5.50(-1)&7.30(-1)&2.32(5)&1.19(-5)&2.30(-1)&2.81(-6)&5.54(-6)&2.39(-6)&1.32(-6)&9.98(-9)&3.05(-5)\\
    \hline
    A$_{\rm v}=$ 3 mag &&&&&\\
    Model 0 &2.1(-2)&1.02(-2)&4.26(-2)&3.14(-1)&2.06(-1)&2.06(-1)&2.81(3)&7.40(-6)&1.23(-1)&1.52(-5)&1.12(-4)&7.34(-5)&4.29(-6)&4.30(-7)&9.56(-4)\\
    Model 8 &5.07(-3)&2.06(-4)&9.33(-3)&7.19(-1)&3.29(-1)&1.05&4.98(3)& 1.15(-6)&5.23(-2)&1.87(-6)&1.44(-4)&6.61(-5)&1.67(-6)&4.56(-7)&5.42(-4)\\
    Model 12 &4.71(-3)&1.48(-4)&7.50(-4)&8.36(-1)&3.52(-1)&1.61&6.96(3)&6.76(-7)&4.19(-2)&1.08(-6)&1.20(-4)&5.06(-5)&1.29(-6)&3.77(-7)&3.55(-4)\\
    Model 19 &1.37(-3)&1.33(-6)&6.20(-3)&1.16(-2)&5.23(-3)&1.22(-1)&7.55(2)&1.81(-6)&3.62(-2)&8.21(-6)&1.54(-5)&6.92(-6)&5.37(-5)&2.91(-5)&2.99(-3)\\
    \hline
    A$_{\rm v}=$ 5 mag &&&&&\\
    Model 0 &1.32(-3)&5.83(-5)&4.10(-3)&4.52(-2)&3.17(-2)&8.68(-2)&2.25(2)&5.89(-6)&9.54(-2)&1.82(-5)&2.01(-4)&1.41(-4)&8.38(-6)&4.55(-6)&7.31(-3)\\
    Model 8 &3.54(-4)&1.34(-5)&5.62(-4)&1.17(-1)&5.71(-2)&5.07(-1)&3.26(2)&1.09(-6)&5.27(-2)&1.72(-6)&3.58(-4)&1.75(-4)&9.56(-7)&2.03(-6)&5.19(-3)\\
    Model 12 &2.15(-4)&7.23(-6)&2.62(-4)&1.19(-1)&5.37(-2)&7.21(-1)&3.18(2)&6.78(-7)&4.61(-2)&8.23(-7)&3.72(-4)&1.69(-4)&2.35(-7)&7.73(-7)&3.88(-3)\\
    Model 19 &6.27(-5)&2.53(-9)&6.60(-4)&1.00(-3)&8.01(-4)&4.27(-2)&5.39(1)&1.16(-6)&5.46(-3)&1.23(-5)&1.86(-5)&1.49(-5)&6.40(-4)&2.13(-3)&2.94(-2)\\
    \hline
    A$_{\rm v}=$ 8 mag &&&&&\\
    Model 0 &7.25(-4)&2.50(-5)&2.83(-3)&2.23(-2)&1.56(-2)&4.75(-2)&1.39(2)&5.19(-6)&7.28(-2)&2.03(-5)&1.60(-4)&1.12(-4)&1.58(-5)&7.68(-6)&8.66(-3)\\
    Model 8 &2.58(-4)&8.83(-6)&5.92(-4)&7.57(-2)&3.64(-2)&3.41(-1)&2.56(2)&1.01(-6)&4.61(-2)&2.31(-6)&2.96(-4)&1.42(-4)&2.60(-6)&4.68(-6)&5.48(-3)\\
    Model 12 &1.71(-4)&5.80(-6)&2.71(-4)&9.20(-2)&4.13(-2)&5.50(-1)&2.61(2)&6.53(-7)&4.33(-2)&1.03(-6)&3.51(-4)&1.58(-4)&7.71(-7)&2.24(-6)&4.68(-3)\\
    Model 19 &3.59(-5)&5.03(-11)&7.78(-5)&2.62(-4)&1.93(-4)&1.63(-2)&3.48(1)&1.03(-6)&7.75(-4)&2.23(-6)&7.52(-6)&5.53(-6)&6.02(-3)&1.68(-2)&3.61(-2)\\
    \hline
    A$_{\rm v}=$ 20 mag &&&&&\\
    Model 0 &6.74(-4)&2.05(-5)&2.63(-3)&2.11(-2)&1.47(-2)&3.61(-2)&1.38(2)&4.88(-6)&6.20(-2)&1.90(-5)&1.55(-4)&1.06(-4)&1.89(-5)&7.20(-6)&8.53(-3)\\
    Model 8 &1.65(-4)&4.10(-6)&6.02(-4)&3.80(-2)&1.76(-2)&1.82(-1)&1.91(2)&8.64(-7)&3.38(-2)&3.14(-6)&1.99(-4)&9.21(-5)&7.12(-6)&9.41(-6)&5.42(-3)\\
    Model 12 &1.16(-4)&2.97(-6)&3.68(-4)&4.83(-2)&2.07(-2)&3.03(-1)&2.12(2)&5.49(-7)&3.17(-2)&1.74(-6)&2.27(-4)&9.77(-5)&3.93(-6)&7.84(-6)&4.63(-3)\\
    Model 19 &3.37(-5)&3.72(-11)&4.15(-5)&1.27(-4)&4.91(-5)&1.11(-2)&3.27(1)&1.03(-6)&6.54(-4)&1.27(-6)&3.88(-6)&1.50(-6)&7.59(-3)&1.63(-3)&3.57(-2)\\
    \hline
    \end{tabular}}
\end{table}
\end{landscape}

\section*{Acknowledgments}

EB would like to thank Prof. David Williams and Dr S. Kaviraj for
useful comments, discussion and help during the preparation and
redaction of this work. EB acknowledges the rolling grants
'Astrophysics at Oxford' PP/EE/E001114/1 and ST/H504862/1 from the
UK Research Councils and the John Fell OUP Research fund, ref
092/267. We also thank the referee for providing very useful
comments which improved significantly the paper. TAB thanks the
Spanish MICINN for funding support through grants AYA2009-07304
and CSD2009-00038. TAB is supported by a CSIC JAE-DOC research
contract.

\newcommand{\apj}[1]{ApJ, }
\newcommand{\apss}[1]{Ap\&SS, }
\newcommand{\aj}[1]{Aj, }
\newcommand{\apjs}[1]{ApJS, }
\newcommand{\apjl}[1]{ApJ Letter, }
\newcommand{\aap}[1]{A\&A, }
\newcommand{\aapr}[1]{A\&A Review, }
\newcommand{\aaps}[1]{A\&A Suppl. Series, }
\newcommand{\araa}[1]{Annu. Rev. A\&A, }
\newcommand{\aaas}[1]{A\&AS, }
\newcommand{\bain}[1]{Bul. of the Astron. Inst. of the Netherland,}
\newcommand{\mnras}[1]{MNRAS, }
\newcommand{\nat}[1]{Nature, }
\newcommand{\araaa}[1]{ARA\&A, }
\newcommand{\planss}[1]{Planet Space Sci., }
\newcommand{\jrasc}[1]{Jr\&sci, }
\newcommand{\pasj}[1]{PASJ, }
\newcommand{\rmxaa}[1]{RMxAA, }

\bibliographystyle{mn2e_2}
\bibliography{references}

\bsp

\label{lastpage}

\end{document}